\documentclass[twocolumn,showpacs,preprintnumbers,amsmath,amssymb,prb]{revtex4}
\usepackage{graphicx}% Include figure files
\usepackage{dcolumn}% Align table columns on decimal point
\usepackage{bm}% bold math
\usepackage{enumerate}

\begin{document}

\preprint{APS/123-QED}

\title{Anisotropic magnetic field responses of ferroelectric polarization in a trigonal multiferroic CuFe$_{1-x}$Al$_x$O$_2$ ($x=0.015$)}

\author{Taro Nakajima}
\email{E-mail address: nakajima@nsmsmac4.ph.kagu.tus.ac.jp}
\author{Setsuo Mitsuda}
\author{Shunsuke Kanetsuki}
\author{Motoyoshi Yamano}
\author{Shunsuke Iwamoto}
\affiliation{Department of Physics, Faculty of Science, Tokyo University of Science, Tokyo 162-8601, Japan}%
\author{Yukihiko Yoshida}
\affiliation{Department of Applied Physics, Faculty of Science, Tokyo University of Science, Tokyo 162-8601, Japan}%
\author{Hiroyuki Mitamura}
\author{Yoshiki Sawai}
\author{Masashi Tokunaga}
\author{Koichi Kindo}
\affiliation{Institute for Solid State Physics, University of Tokyo, Kashiwa 277-8581, Japan}
\author{Karel Proke\v{s}}
\author{Andrey Podlesnyak}
\affiliation{Helmholtz-Centre Berlin for Materials and Energy, SF-2, Glienicker Stra\ss e 100, Berlin 14109, Germany}%

\begin{abstract}
We have investigated magnetic field dependences of a ferroelectric incommensurate-helimagnetic order in a trigonal magneto-electric (ME) multiferroic %
CuFe$_{1-x}$Al$_x$O$_2$ with $x=0.015$, which exhibits the ferroelectric phase as a ground state, %
by means of neutron diffraction, magnetization and dielectric polarization measurements under magnetic fields applied along various directions. %
From the present results, we have established the $H$-$T$ magnetic phase diagrams for the three principal directions of magnetic fields; %
(i) parallel to the $c$ axis, (ii) parallel to the helical axis, and (iii) perpendicular to the $c$ and the helical axes. %
While the previous dielectric polarization ($P$) measurements on CuFe$_{1-x}$Ga$_x$O$_2$ with $x=0.035$ have demonstrated that %
the magnetic field dependence of the `magnetic domain structure' results in distinct magnetic field responses of $P$ %
[S. Seki \textit{et al.}, Phys. Rev. Lett., {\bf 103} 237601 (2009)], % 
the present study have revealed that the anisotropic magnetic field dependence of the ferroelectric helimagnetic order %
`in each magnetic domain' %
can be also a source of a variety of magnetic field responses of $P$ in CuFe$_{1-x}A_x$O$_2$ systems ($A=$Al, Ga). %
\end{abstract}

\maketitle

\section{INTRODUCTION}

Nonlinear magneto-electric (ME) effects, in particular magnetic-field control of ferroelectric polarization ($P$), %
in ME-multiferroics have been intensively studied since the discovery of the colossal ME-effect in some magnetically frustrated transition metal oxides.\cite{Kimura_nature,Hur_TM2O5,PRL_Ni3V2O8,PRL_MnWO4,Yamasaki_PRL_conical} %
In most of the ME-multiferroics, the magnetic field dependences of $P$ have been attributed to changes in the %
magnetic structures. %
For example, the magnetic-field-induced 90$^{\circ}$-flop of $P$ in TbMnO$_3$ has been explained by the %
first-order magnetic phase transition from the $bc$-plane cycloidal magnetic ordering to the $ac$-plane cycloidal %
magnetic ordering.\cite{TbMnO3_Pa,TbGdMnO3_Pa} %
On the other hand, the recent experimental works have pointed out that magnetic control of `magnetic domain structures' %
can be an another ingredient for the ME-effects, in some relatively high-symmetry (trigonal, tetragonal or hexagonal) multiferroics.\cite{Kimura_CuCrO2,Seki_CuCrO2,Murakawa_ZnCr2Se4,Seki_CFGO} %
For example, Kimura \textit{et al.} have argued that an anisotropic magnetic field dependence of $P$ in a trigonal %
ME-multiferroic CuCrO$_2$ can be ascribed to the magnetic field dependence of the volume fractions of the three magnetic domains, %
in which the three magnetic modulation wave vectors characterizing these domains %
are equivalent to each other because of the trigonal symmetry of the crystal structure.\cite{Kimura_CuCrO2} %
This suggests that trigonal (or tetragonal, hexagonal) ME-multiferroics can provide opportunities for realizing %
a variety of magnetic field responses of $P$. %

\begin{figure}[t]
\begin{center}
	\includegraphics[clip,keepaspectratio,width=7.5cm]{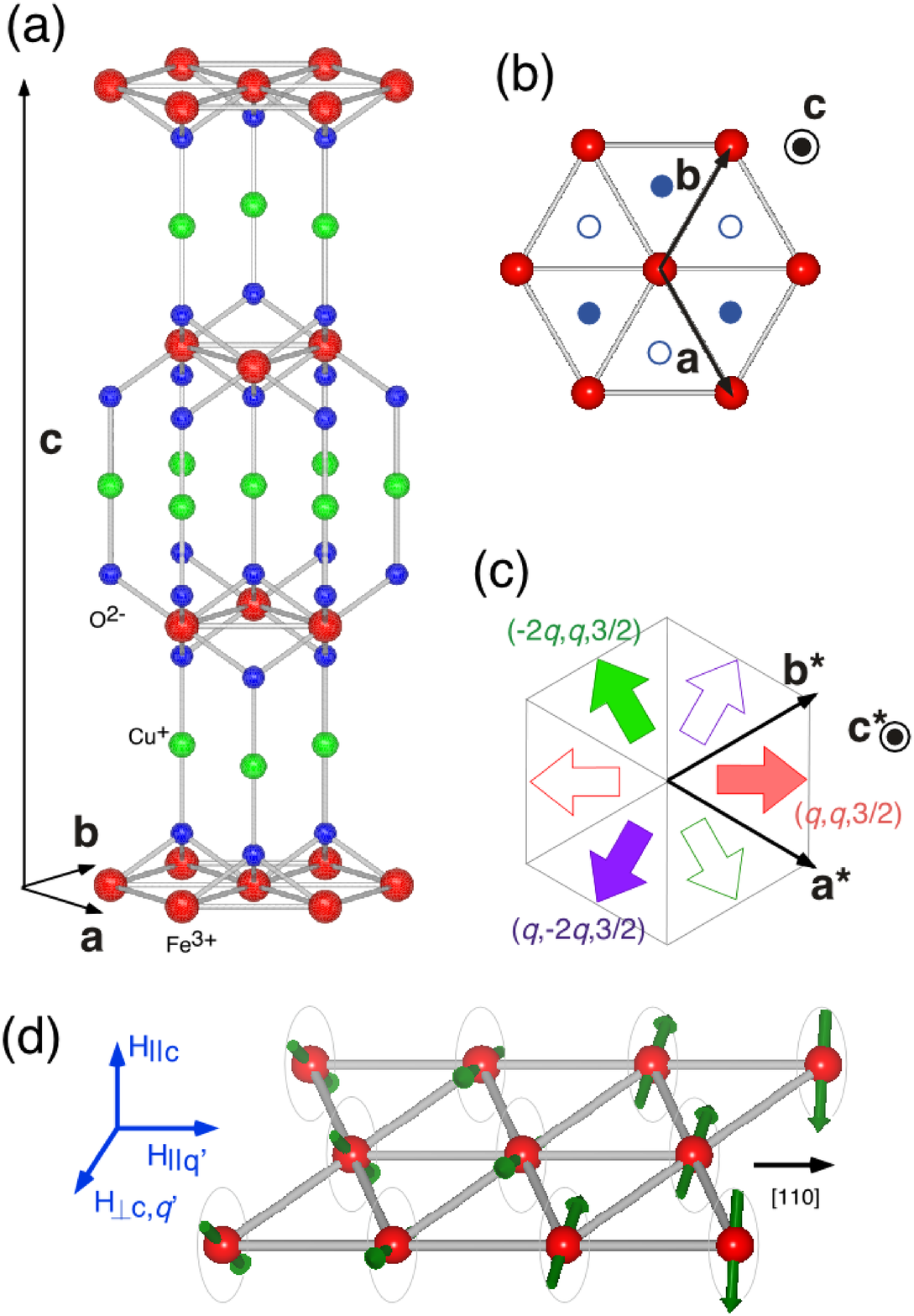}
	\caption{(a) Crystal structure of CuFeO$_2$. (b) The definitions of the hexagonal basis and the arrangements of the O$^{2-}$ ions %
	above (open blue circles) and below (filled blue circles) a Fe$^{3+}$ triangular lattice layer. %
	(c) Schematic drawing of the $q'$-vectors of the three $q$-domains and the reciprocal lattice basis. %
	(d) Illustration of the magnetic structure in the FE-ICM phase and the directions of $H_{\parallel c}$, $H_{\parallel q'}$ and $H_{\perp c,q'}$. %
	}
	\label{cryst_str}
\end{center}
\end{figure}

In the past several years, a trigonal ME-multiferroic CuFeO$_2$ (CFO) has been the subject of increasing interest %
as a ME-multiferroic because of the discovery of the ferroelectricity in a magnetic-field-induced phase.\cite{Kimura_CuFeO2} % 
Subsequent studies have elucidated that %
the ferroelectric phase is stabilized even under zero magnetic field by substituting small amount of nonmagnetic Al$^{3+}$ or Ga$^{3+}$ ions %
for the magnetic Fe$^{3+}$ sites.\cite{Kanetsuki_JPCM,Seki_PRB_2007,Terada_CFGO} % 
The magnetic structure in the ferroelectric phase has been determined to be a proper-screw-type helical magnetic structure.\cite{SpinNoncollinearlity} %
The magnetic modulation wave vector is $(q,q,\frac{3}{2})$ where $q=0.202\sim0.210$, and the helical axis is parallel to the $[110]$ axis.\cite{SpinNoncollinearlity,CompHelicity} %
Hereafter, we refer to the ferroelectric phase as ferroelectric incommensurate-magnetic (FE-ICM) phase. %
Recent polarized neutron diffraction studies have revealed  that the spin-helicity, left-handed (LH) or right-handed (RH) helical arrangement of spins, %
determines the polarity of the local ferroelectric polarization emerging along the helical axis.\cite{CFAO_Helicity,CompHelicity} %
Because of the threefold rotational symmetry about the hexagonal $c$ axis (see Fig.\ref{cryst_str}(a)), %
the magnetic ordering  with the wave vector of $(q,q,\frac{3}{2})$ results in three magnetic domains %
whose wave vectors of $(q,q,\frac{3}{2})$, $(q,-2q,\frac{3}{2})$ and $(-2q,q,\frac{3}{2})$ are crystallographically equivalent to each other, %
as illustrated in Fig. \ref{cryst_str}(c). %
In this paper, we refer to these three domains as `$q$-domains'. %

Quite recently, Seki \textit{et al.} have reported that the `magnetic domain structure' in the FE-ICM phase, specifically the volume fractions of the three $q$-domains, can be controlled by %
applying a magnetic field in the triangular lattice plane.\cite{Seki_CFGO} %
They performed magnetization and dielectric polarization measurements on CuFe$_{1-x}$Ga$_x$O$_2$ (CFGO) with $x=0.035$, %
in which the FE-ICM phase shows up as a ground state, under various magnitudes and directions of magnetic fields. %
These measurements revealed that the `in-plane'-magnetic-field-dependence of the volume fractions of the $q$-domains results in distinct %
magnetic field responses of $P$, for example, 120$^{\circ}$-flop of $P$ by the magnetic fields rotating in the $c$-plane. %
However, magnetic field dependence of the magnetic ordering `in each $q$-domain' was not discussed, %
because it is quite difficult to extract the information on the magnetic ordering in each $q$-domain from the results of the macroscopic polarization or magnetization measurements. %
In order to completely understand magnetic field dependences of $P$ in slightly diluted CFO systems, therefore, %
it is critical to elucidate both of the magnetic field dependences of the `magnetic ordering in each $q$-domain' and %
the `magnetic domain structure'. %

In the present study, we have investigated magnetic field dependences of $P$ and the FE-ICM order %
in CuFe$_{1-x}$Al$_x$O$_2$ (CFAO) with $x=0.015$, which exhibits the FE-ICM phase as a ground state, %
by means of neutron diffraction, magnetization and dielectric polarization measurements under various directions of magnetic fields. %
In order to elucidate the magnetic field dependence of the FE-ICM order `in a $q$-domain', we have established $H$-$T$ magnetic phase %
diagrams for the three principal directions of magnetic fields, specifically, (i) parallel to the $c$ axis, (ii) parallel to %
the helical axis, and (iii) perpendicular to the $c$ and the helical axes. %
The present results have revealed that the anisotropic magnetic field dependence of the FE-ICM order `in each $q$-domain' %
results in a variety of magnetic field responses of $P$ as well as the magnetic field dependence of the `magnetic %
domain structure' does. %

We have also investigated a magnetic field dependence of sensitivity of $P$ to a poling electric field ($E_p$). %
In our previous polarized neutron diffraction and in-situ pyroelectric measurements on CFAO($x=0.015$) and CFGO($x=0.035$) in zero field, %
we have found that the Al-substitution more significantly reduced the sensitivity of $P$ to $E_p$ as compared to the Ga-substitution.\cite{CompHelicity} %
These measurements have also revealed that CFGO($x=0.035$) exhibits a homogeneous FE-ICM state, that is relatively close to %
the long-range-ordered state, in contrast to CFAO($x=0.015$), that has an inhomogeneous domain state in zero field.\cite{CompHelicity} %
We have thus concluded that the `inhomogeneity' of the FE-ICM order, which must be relevant to the mobility of the %
magnetic domain walls, determines the sensitivity of $P$ to $E_p$. %
On the other hand, our previous neutron diffraction measurements under applied magnetic fields %
showed that the homogeneous FE-ICM state can be realized even in CFAO($x=0.015$) by applying a magnetic field along the $c$ axis.\cite{SpinNoncollinearlity} %
Therefore, the sensitivity of $P$ to $E_p$ in CFAO($x=0.015$) is  expected to be controlled by an application of a magnetic field. %
This is the reason why we have selected CFAO($x=0.015$) sample for the present study. %

This paper is organized as follows. In Sec. \ref{Experimental_details}, we describe experimental details. %
Sec. \ref{Sec_RESULT} consists of two subsections. %
In Sec. \ref{Sec_PhaseDiagram}, we present anisotropic $H$-$T$ magnetic phase diagrams for a $q$-domain %
using the results of the neutron diffraction measurements under the three directions of the magnetic fields, and also present %
the results of pyroelectric measurements under steady magnetic fields. %
In Sec. \ref{Sec_MEeffect}, we present magnetic field variations of $P$ and the FE-ICM order using the results of neutron diffraction, %
magnetization and dielectric polarization measurements under magnetic fields. %
In Sec. \ref{Conc}, we summarize our conclusions. %

\section{EXPERIMENTAL DETAILS}
\label{Experimental_details}

Single crystal of CuFe$_{1-x}$Al$_x$O$_2$ with $x=0.015$ was %
prepared by the floating zone technique \cite{Zhao_FZ}. %
For the measurements of $P$, %
the sample was cut into a thin plate ($\sim 2\times 4\times 0.1$ mm$^3$). %
Silver paste was applied on the $[1\bar{1}0]$ surface of the sample to form the electrodes. %
In the pyroelectric measurements under steady magnetic fields, %
the pyroelectric current was measured under zero electric field with increasing temperature, using an electrometer (Keithley 6517A). %
Before each pyroelectric measurement, we performed cooling %
with an applied poling electric field ($E_p$) from 15 K to 2 K. %
After the poling electric field was removed at 2 K, the sample was allowed to discharge for about 40 minutes in order to reduce residual current. %
The typical magnitude of $E_p$ in the present pyroelectric measurements was $\sim 250$ kV/m. %
We also investigated the $E_p$ dependence of $P$ up to $E_p=2.0$ MV/m. %
External magnetic fields up to 5 T were provided by the Magnetic Property Measurement System (Quantum Design Inc.). %

We have performed dielectric polarization and magnetization measurements on CFAO($x=0.015$) under pulsed magnetic fields. %
The pulsed high magnetic fields up to 55 T were generated by a non-destructive magnet in the International MegaGauss %
Science Laboratory in ISSP, the University of Tokyo. %
The magnetization along the field direction was measured by the induction method using coaxial pick-up coils. %
The dimensions of the single crystal sample used for the magnetization measurements were $1.8\times 1.8 \times 2.3$ mm$^{3}$. %
Following the pioneer work on the dielectric polarization measurement under pulsed magnetic fields by Mitamura \textit{et al.},\cite{Mitamura} %
we detected the $H$-induced change in $P$ by monitoring the polarization current through a voltage drop in the %
shunt resistance connected in series. %
By integrating the current with respect to time, we obtained the magnetic field variations of $P$. %
In this measurement, the poling electric fields ($E_p$) were continuously applied during all the measurements of the $P$-$H$ curve. %
The magnitude of $E_p$ was typically $\sim$200 kV/m. %
The single crystal sample for this measurement was identical to the sample used for the present pyroelectric measurements. %

The neutron diffraction measurements under applied field were carried out at the two-axis neutron diffractometer %
E4 installed at Berlin Neutron Scattering Center (BENSC) in Helmholtz Centre Berlin for Materials and Energy. %
A typical dimension of the single crystal samples for the neutron experiments was $\sim 3\times 4\times 4.5$ mm$^3$. %
External magnetic fields along the hexagonal [001], [110], and $[1\bar{1}0]$ directions were provided by %
the cryomagnets HM-1, HM-2 and VM-1, whose maximum fields are 6T, 4T, and 14.5T, respectively. %
Note that the magnetic field along [001] axis was canted by $\sim 12^{\circ}$ from the [001] axis toward [110] axis, %
because of limitations of the windows of the horizontal field cryomagnet HM-1. 
For all the neutron diffraction measurements, the sample was mounted in the cryomagnet with a %
($h,h,l$) scattering plane. % 
In the experiments with the [001] and [1$\bar{1}$0] magnetic fields, the collimation was 40'-40'-40', and %
a single detector was used. %
In the experiment with the [110] magnetic field, the collimation was 40'-40'-open, and a 2D-position-sensitive-detector %
was used. %

In this paper, we have employed a conventional hexagonal basis defined as shown in Fig. \ref{cryst_str}(a), %
in order to describe the directions of the magnetic propagation vectors of the three $q$-domains, %
although structural transitions from the original trigonal structure to a monoclinic structure have been reported for %
some of the magnetically ordered phases (including the FE-ICM phase) of CuFeO$_2$\cite{Terada_CuFeO2_Xray,Ye_CuFeO2,Terada_14.5T} and CuFe$_{1-x}$Al$_x$O$_2$\cite{CFAO_2q}. %

\begin{figure*}[t]
\begin{center}
	\includegraphics[clip,keepaspectratio,width=0.95\textwidth]{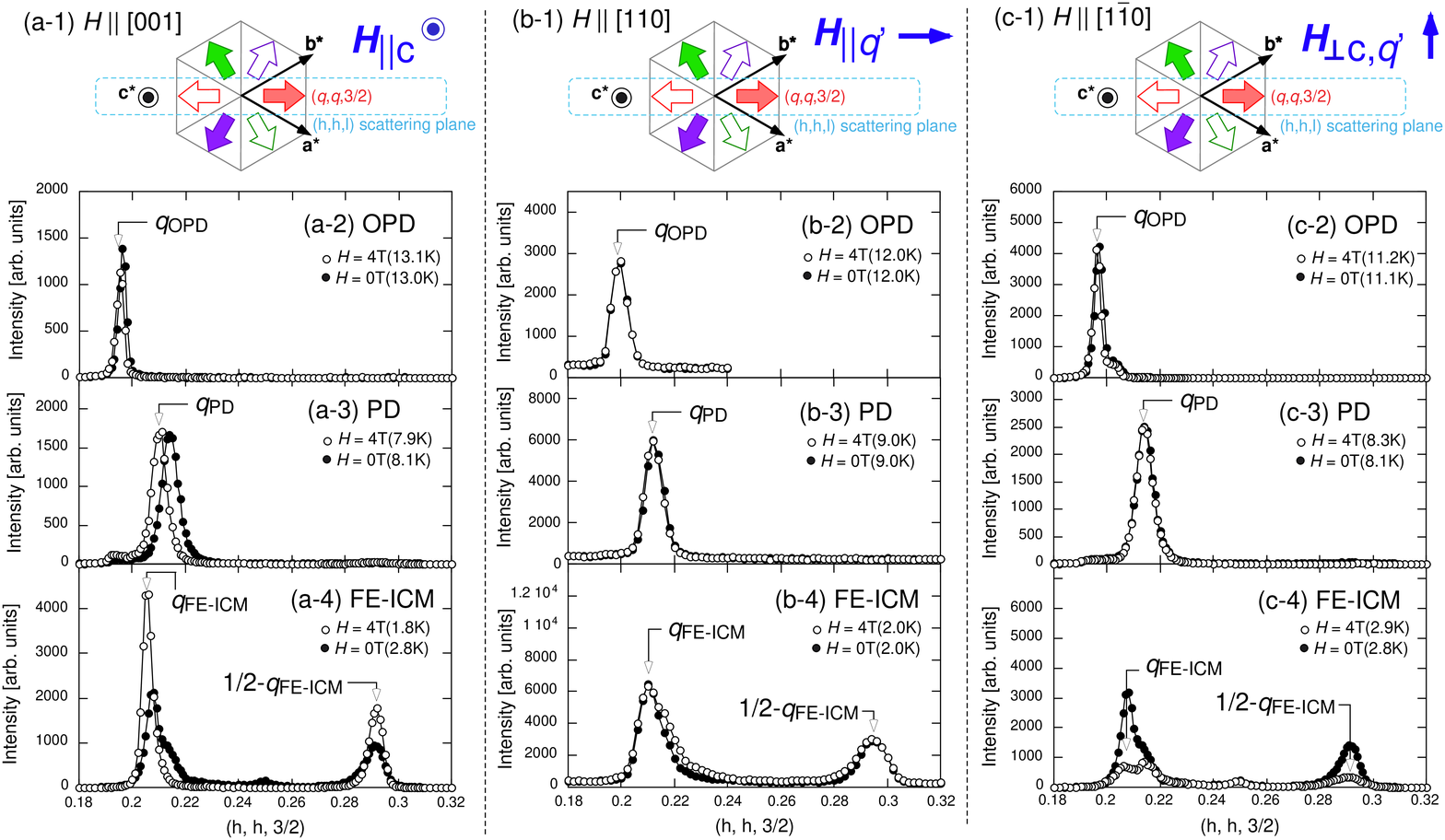}
	\caption{[(a-1),(b-1),(c-1)] The relationships between the applied magnetic fields and $q'$-vector of the $q$-domain observed %
	in the present neutron diffraction measurements under (a-1) $H_{\parallel c}$, (b-1) $H_{\parallel q'}$ and (c-1) $H_{\perp c,q'}$. %
	The magnetic diffraction profiles measured in the cooling processes under [(a-2)-(a-4)] $H_{\parallel c}$, %
	[(b-2)-(b-4)] $H_{\parallel q'}$ and [(c-2)-(c-4)] $H_{\perp c,q'}$. %
	The data observed in the magnetic fields are shown by the open circles, and the those in zero field are shown by the black filled circles. %
	Note that the difference in the width of the magnetic diffraction profiles between the measurements under $H_{\parallel q'}$ and %
	the others is due to the different experimental configurations, specifically the collimations and %
	the detectors. See Sec.\ref{Experimental_details} for details. %
	}
	\label{FC_Neutron}
\end{center}
\end{figure*}

\section{RESULTS AND DISCUSSIONS}
\label{Sec_RESULT}
\subsection{Magnetic field dependence of the phase transitions in a $q$-domain}
\label{Sec_PhaseDiagram}

\subsubsection{Neutron diffraction measurements under steady magnetic fields}

We firstly investigated the temperature variations of the magnetic ordering in a $q$-domain under steady magnetic fields, by means of the %
neutron diffraction measurements. %
In order to define the relationship between the directions of the magnetic fields and the magnetic structure in a $q$-domain, %
we introduce the $c$-plane-projection of the $q$-vector, $q'=(q,q,0)$. %
In the present neutron diffraction measurements, %
we applied magnetic fields along the three directions; (i) nearly parallel to the [001] axis,\cite{comment2} (ii) parallel to the [110] axis, and %
(iii) parallel to the $[1\bar{1}0]$ axis. %
Since the magnetic reflections on the $(h,h,l)$ scattering plane belong to the $q$-domains with the wave vector of $(q,q,\frac{3}{2})$, %
the directions of the [110] axis corresponds to the $q'$-vector of the $(q,q,\frac{3}{2})$-domain. %
Hereafter, we refer to the magnetic fields along these three directions as $H_{\parallel \rm c}$, $H_{\parallel q'}$ and $H_{\perp c,q'}$ %
(see Fig. \ref{cryst_str}(d)). %

Before discussing the present results, %
we should review the magnetic phase transitions in CFAO ($x=0.015$) in zero magnetic field. %
As reported by Terada \textit{et al.},\cite{Terada_x_T} CFAO($x=0.015$) has three magnetically ordered phases in zero magnetic field. %
The typical magnetic diffraction profiles of the $(h,h,\frac{3}{2})$ reciprocal lattice scans in zero field are %
shown in Fig. \ref{FC_Neutron} by black filled circles. %
With decreasing temperature from the paramagnetic (PM) phase, %
two collinear-incommensurate magnetic phases show up; the higher temperature phase is %
the oblique-partially-disorderd (OPD) phase and the lower temperature phase is the partially disordered (PD) phase. %
Both of the magnetic reflections corresponding to the OPD and PD magnetic orderings are assigned as $(q,q,\frac{3}{2})$. %
The incommensurate wave number for the OPD phase, $q_{\rm OPD}\sim 0.195$ is almost independent of temperature, %
while that for the PD phase, $q_{\rm PD}$, varies from 0.20 to 0.22 with decreasing temperature. %
Previous neutron diffraction measurements by Terada \textit{et al.}\cite{Terada_x_T,Terada_FONDER} have revealed that the magnetic moments %
in the PD  and OPD phases are canted by about 12$^{\circ}$, 50$^{\circ}$ from the $[001]$ direction toward the $[1\bar{1}0]$ direction, %
respectively. %
The ground state of CFAO ($x=0.015$) is the FE-ICM phase, as mentioned in introduction. %
The magnetic diffraction profile in the FE-ICM phase is %
characterized by the two magnetic reflections assigned as $(q,q,\frac{3}{2})$ and $(\frac{1}{2}-q,\frac{1}{2}-q,\frac{3}{2})$\cite{comment1} using the %
hexagonal basis, as shown in Fig. \ref{FC_Neutron}(a-3). %
The magnetic modulation wave number in the FE-ICM phase, $q_{\rm FE-ICM}=0.202\sim 0.210$, slightly depends on the Al-concentration %
and applied magnetic fields, but is almost independent of temperature. %
It should be noted that a small peak at $(\frac{1}{4},\frac{1}{4},\frac{3}{2})$ %
corresponds to the collinear-commensurate 4-sublattice phase, which coexists with the FE-ICM phase at low temperatures because of a %
slight macroscopic inhomogeneity of the Al-concentration in the samples.\cite{Terada_x_T} %

We now discuss the results of the present neutron diffraction measurements under applied magnetic fields. %
In Fig. \ref{FC_Neutron}, we show the magnetic diffraction profiles measured on cooling %
under $H_{\parallel c}$, $H_{\parallel q'}$ and $H_{\perp c,q'}$ of 4 T. %
In all the three cooling, the successive magnetic transitions [OPD$\rightarrow$PD$\rightarrow$FE-ICM] were observed, %
and no strong magnetic field dependences of the magnetic diffraction profiles were found in the OPD and PD phase, %
although the wave number of the %
PD phase is slightly dependent on $H_{\parallel c}$. %
However, we found that the magnetic diffraction profiles in the FE-ICM phase under $H_{\parallel c}$, $H_{\parallel q'}$ and $H_{\perp c,q'}$ %
are remarkably different from each other, %
as shown in Figs. \ref{FC_Neutron}(a-3), \ref{FC_Neutron}(b-3) and \ref{FC_Neutron}(c-3). %

Figure \ref{FC_Neutron}(b-3) shows the magnetic diffraction profile in the FE-ICM phase under $H_{\parallel q'}$, %
suggesting that the magnetic field applied along the helical axis does not result in the drastic change in the FE-ICM order. %
In contrast, the magnetic fields applied perpendicular to the helical axis, $H_{\parallel c}$ and $H_{\perp c,q'}$, %
significantly affect the magnetic diffraction profiles in the FE-ICM phase. %
Figure \ref{FC_Neutron}(a-3) shows that $H_{\parallel c}$ sharpens the magnetic diffraction profile in the FE-ICM phase, %
as was partly reported in Ref.\onlinecite{SpinNoncollinearlity}. %
This suggests that the FE-ICM order under $H_{\parallel c}$ is relatively close to the long-range-ordered state, while %
the FE-ICM order is inhomogeneous domain state in zero magnetic field,\cite{CompHelicity} as mentioned in introduction. %
This $H_{\parallel c}$-dependence of the magnetic correlation might be attributed to the local lattice distortions due to the %
Al-substitution. %

The previous studies on CFAO\cite{CompHelicity} and CFGO\cite{Terada_CFGO,CompHelicity} have pointed out that the Al-substitution should result in local lattice distortions because %
the ionic radius of Al$^{3+}$ is much smaller than that of Fe$^{3+}$, and have also suggested that %
this local lattice distortion strongly disturbs a coherent magnetic ordering in the FE-ICM phase %
because the distortions randomly lift the local degeneracy in the competing exchange interactions. % 
On the other hand, the previous synchrotron radiation x-ray diffraction measurements on pure CFO under applied magnetic fields %
revealed that the lattice constants in the FE-ICM phase vary with $H_{\parallel c}$;\cite{Terada_14.5T,Terada_CFO_Pulse} %
specifically, the lattice constant along the $[110]$ direction, which is the $b$ axis in the monoclinic notation, %
linearly decreases with increasing $H_{\parallel c}$. %
Taking account of these results, one can expect that the local lattice distortion relaxes with increasing $H_{\parallel c}$. %
For further investigation on this problem, x-ray diffraction measurements on CFAO system under applied magnetic fields are required. %

\begin{figure}[t]
\begin{center}
	\includegraphics[clip,keepaspectratio,width=7.5cm]{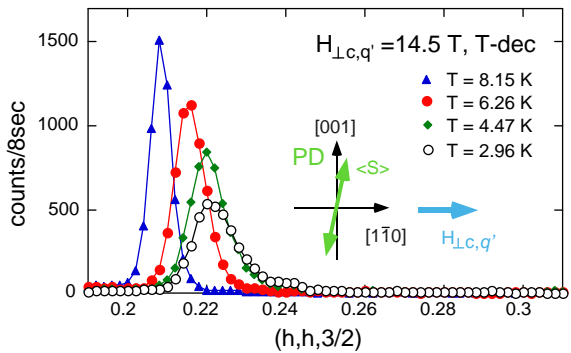}
	\caption{The temperature variation of the magnetic diffraction profile of the $(h,h,\frac{3}{2})$ reciprocal lattice scans under $H_{\perp c,q'}$ of 14.5 T. %
	Inset shows relationship between the directions of $H_{\perp c,q'}$ and the magnetic moments of the PD magnetic order in zero field. }
	\label{FC_14.5T}
\end{center}
\end{figure}

\begin{figure}[t]
\begin{center}
	\includegraphics[clip,keepaspectratio,width=7.5cm]{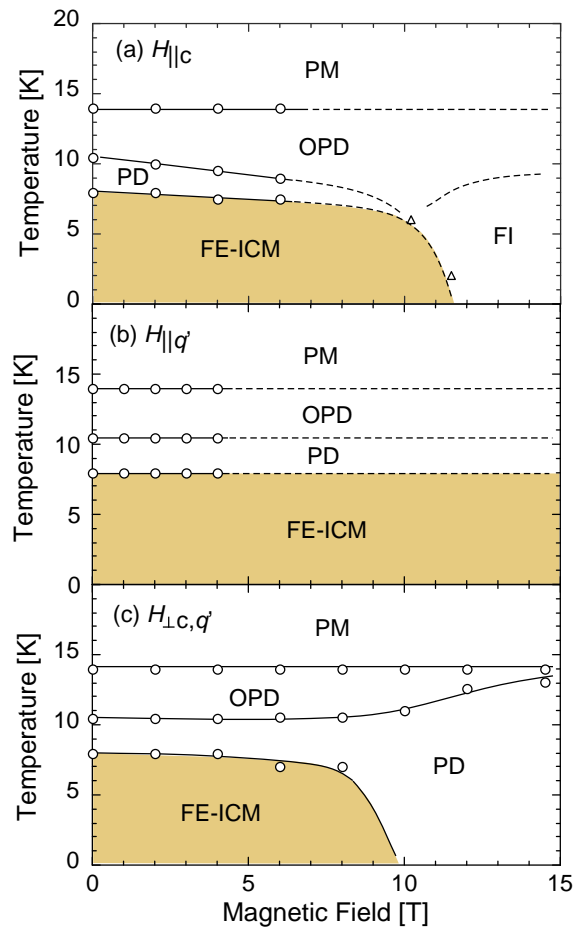}
	\caption{[(a)-(c)] The $H$-$T$ magnetic phase diagrams for (a) $H_{\parallel c}$, (b) $H_{\parallel q'}$ and (c) $H_{\perp c,q'}$. %
	Open circles and solid lines denote the phase boundaries determined by the present neutron diffraction measurements on cooling. %
	The phase boundaries in the high field region of the $H_{\parallel c}$-$T$ phase diagram (dashed lines) were drawn %
	in analogy of the $H_{\parallel c}$-$T$ phase diagram of CFAO($x=0.02$).\cite{Terada_CFAO_H-T} %
	To check the reasonability of the phase boundaries in (a), we have shown the transition field between the FE-ICM phase %
	and the FI phase determined by the dielectric polarization measurements %
	under pulsed magnetic fields presented in Sec. \ref{Pulse} (open triangles). %
	}
	\label{FC_PhaseDiagram}
\end{center}
\end{figure}

In contrast to $H_{\parallel c}$, in the cooling process under $H_{\perp c,q'}$, %
the magnetic diffraction corresponding to the FE-ICM order is rather diffusive. %
This implies that the FE-ICM order is suppressed by $H_{\perp c,q'}$. %
Actually, as shown in Fig. \ref{FC_14.5T}, the PD to FE-ICM magnetic phase transition is not detected in the cooling process under $H_{\perp q',c} = 14.5$ T, %
and instead, the PD magnetic ordering survives even at 2.9K. %
This suggests that $H_{\perp c,q'}$ favors the %
PD magnetic ordering, whose magnetic moments lie nearly perpendicular to the magnetic field, rather than the proper-screw-type %
magnetic order in the FE-ICM phase. %

\subsubsection{$H$-$T$ magnetic phase diagram for $H_{\parallel c}$, $H_{\parallel q'}$ and $H_{\perp c,q'}$}

In Figs. \ref{FC_PhaseDiagram}(a)-\ref{FC_PhaseDiagram}(c), we now present the $H$-$T$ magnetic phase diagrams for $H_{\parallel c}$, $H_{\parallel q'}$ and $H_{\perp c,q'}$ %
deduced from the present results. %
Although the present neutron %
diffraction measurements could not reach the high field region of the $H_{\parallel q'}$-$T$ phase diagram, %
we have confirmed that the FE-ICM phase extends to the high field region %
by the present dielectric polarization measurements under pulsed magnetic fields up to 30 T %
to be mentioned in Sec. \ref{Pulse}. %
Here, we should emphasize that these phase diagrams represent the magnetic ordering `in a $q$-domain', %
and moreover, those revealed that the FE-ICM order in a $q$-domain %
shows the anisotropic responses for the in-plane magnetic fields of $H_{\parallel q'}$ and $H_{\perp c,q'}$, %
which was not directly observed in the previous macroscopic polarization and magnetization measurements.\cite{Seki_CFGO} %

\begin{figure*}[t]
\begin{center}
	\includegraphics[clip,keepaspectratio,width=0.95\textwidth]{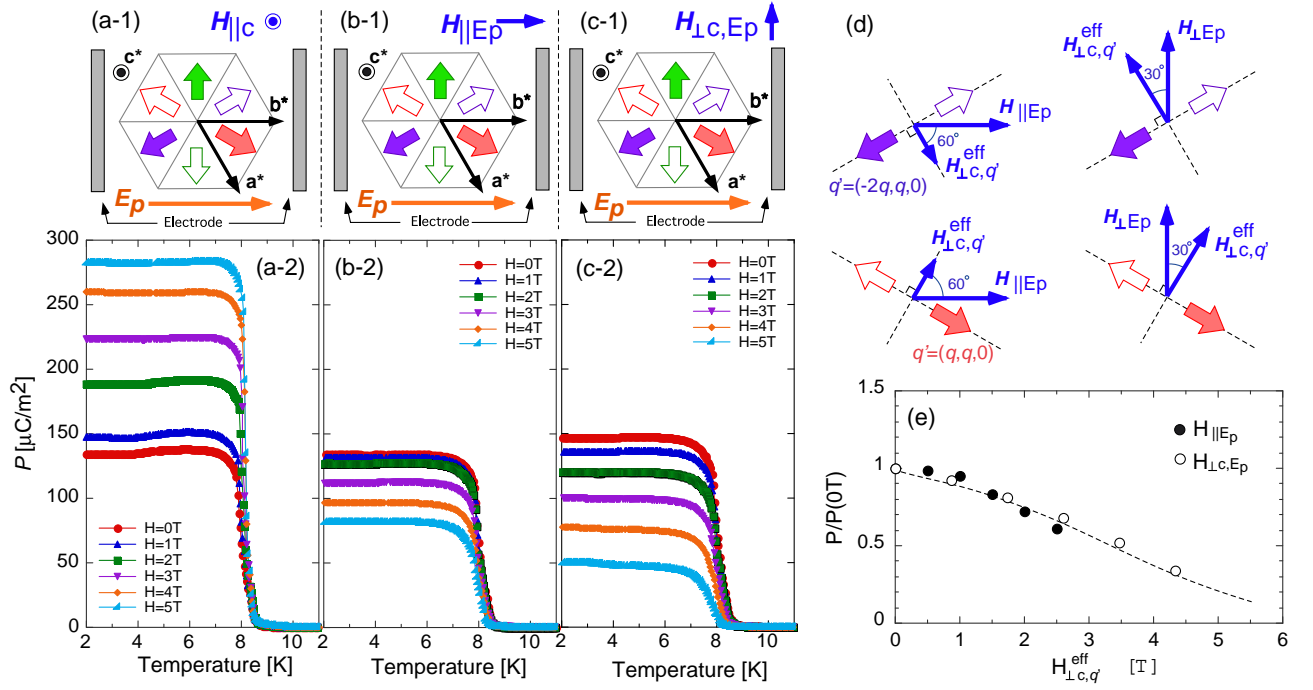}
	\caption{[(a-1),(b-1),(c-1)] The relationships between the applied magnetic fields and $q'$-vectors for (a-1) $H_{\parallel c}$, %
	(b-1) $H_{\parallel E_p}$ and (c-1) $H_{\perp c,E_p}$. %
	The magnetic field dependence of the temperature variations of $P$ under (a-2) $H_{\parallel c}$, %
	(b-2) $H_{\parallel E_p}$ and (c-2) $H_{\perp c,E_p}$. %
	(d) The definition of the $H^{eff}_{\perp c,E_p}$ for the in-plane magnetic fields of $H_{\parallel E_p}$ and $H_{\perp c,E_p}$. %
	(e) The $H^{eff}_{\perp c,E_p}$-dependences of $P$ at 2.0 K normalized to the value of $P$ in zero field. %
	The dashed line is the guide to eyes. %
	}
	\label{FC_Pyro}
\end{center}
\end{figure*}

\subsubsection{Pyroelectric measurements in steady magnetic fields}
\label{Sec_Pyro}

We have also performed pyroelectric measurements under applied magnetic fields, %
to observe the magnetic field dependences of $P$ corresponding to the FE-ICM order observed in the present %
neutron diffraction measurements. %
As mentioned in introduction, the direction of the spontaneous electric polarization in the FE-ICM order is parallel to the %
helical axis,\cite{CFAO_Helicity} that is, the direction of the $q'$-vector. %
In addition, the previous polarized neutron diffraction measurement under applied electric field revealed that an application of a %
poling electric field cannot result in a single $q$-domain state.\cite{CFAO_Helicity,CompHelicity} %
Therefore, the measured electric polarization has to be the sum of the contributions from the domains with the different $q$-vectors. % 
This situation prevent us from investigating the anisotropic ME-responses in a $q$-domain by pyroelectric measurements. %

To overcome this problem, we have thus selected the [120] plane, which is crystallographically equivalent to the $[1\bar{1}0]$ and $[\bar{2}\bar{1}0]$ planes, %
as the electrode surfaces, %
and applied magnetic fields along three principal directions; (i) parallel to the $c$ axis, (ii) parallel to the poling electric field ($E_p$), %
and (iii) perpendicular to the $c$ axis and $E_p$. % 
We refer to these three directions of the magnetic fields as $H_{\parallel c}$, $H_{\parallel E_p}$ and $H_{\perp c,E_p}$, respectively. %
The directions of the poling electric field, the $q'$-vectors of the three domains and the applied magnetic fields are schematically drawn in %
Figs. \ref{FC_Pyro}(a-1), \ref{FC_Pyro}(b-1) and \ref{FC_Pyro}(c-1). %
In this configuration of the electrodes, only the two $q$-domains with the wave vectors of $(q,q,\frac{3}{2})$ and $(q,-2q,\frac{3}{2})$ contribute to %
the measured electric polarization, because the electric polarization vector in the $(-2q,q,\frac{3}{2})$-domain is perpendicular to %
the normal vector of the electrode surfaces. %
In addition, the $H_{\parallel E_p}$- (or $H_{\perp c,E_p}$-) dependence of the FE-ICM order in the $(q,q,\frac{3}{2})$-domains %
is expected to be the same as that in the $(q,-2q,\frac{3}{2})$-domains, because of the symmetry of the crystal and magnetic structures. %
By this configuration of the electrodes, %
the anisotropic in-plane-field-dependences of $P$ can be observed. %

Figures \ref{FC_Pyro}(b-2) and \ref{FC_Pyro}(c-2) show the temperature variations of $P$ %
under $H_{\parallel E_p}$ and $H_{\perp c,E_p}$. %
For both of $H_{\parallel E_p}$ and $H_{\perp c,E_p}$, the magnitude of $P$ decreased with increasing magnetic field. %
Taking account the present neutron diffraction measurements revealing that the FE-ICM order was significantly affected by $H_{\perp c,q'}$ %
and was less affected by $H_{\parallel q'}$, %
we conclude that the $H_{\perp c,q'}$-components of the magnetic fields are relevant to the reduction of $P$. %
Actually, $H_{\perp c,E_p}$ more remarkably reduced $P$ than $H_{\parallel E_p}$. %
We thus show, in Fig. \ref{FC_Pyro}(e), the values of $P$ at $T=2$ K normalized to the values in zero magnetic field as a function of the effective magnetic field %
applied perpendicular to the $c$ axis and the $q'$-vector, $H^{eff}_{\perp c,q'}$ (see Fig. \ref{FC_Pyro}(d)), specifically %
\begin{eqnarray}
H^{eff}_{\perp c,q'}=H_{\perp c,E_p}\cos 30^{\circ}=H_{\parallel E_p}\sin30^{\circ}.
\end{eqnarray}
We found that the $H_{\perp c,E_p}$- and $H_{\parallel E_p}$-dependences of $P$ at $T=2.0$ K are scaled %
by $H^{eff}_{\perp c,q'}$. %
This clearly shows that $H_{\perp c,q'}$ dominates the `in-plane' magnetic field dependence of $P$, %
as was expected from the results of the present neutron diffraction measurements. %
 
\begin{figure}[t]
\begin{center}
	\includegraphics[clip,keepaspectratio,width=7.5cm]{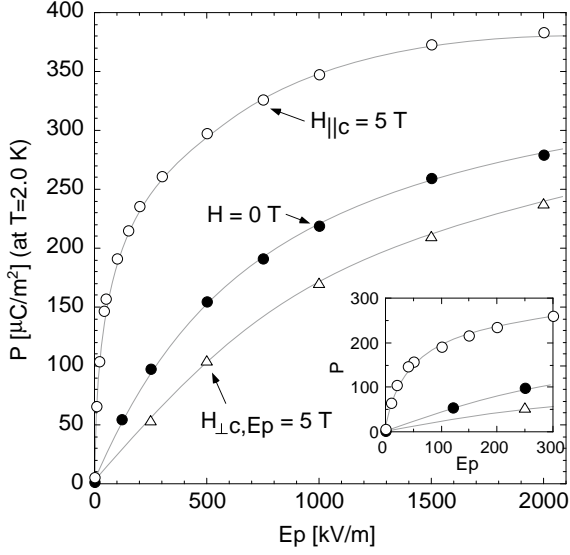}
	\caption{The $E_p$-dependences of the values of $P$ at $T=2.0$ K in zero field, under $H_{\parallel c}=5.0$ T and %
	$H_{\perp c,E_p}=5.0$ T. The inset shows the magnification of the region of $0<E_p<300$ kV/m. %
	The solid gray lines are guides to eyes. %
	}
	\label{E_vs_P}
\end{center}
\end{figure}

As for the origin of the reduction of $P$ under a magnetic field having the $H_{\perp c,q'}$-component, we can propose three possibilities; %
the first one is that the volume fraction of the FE-ICM order in each $q$-domains was reduced because of the PD magnetic order %
retained by the magnetic field, the second is that the magnetic structure of the FE-ICM order was partly modified by the magnetic field, %
and the third is that the sensitivity of $P$ to $E_p$ was reduced because of the reduction of the coherence of the %
magnetic order in the FE-ICM phase under the magnetic field. %
A combination of two or three of them is also possible. %
Although we cannot identify the origin of the reduction of $P$ only from the present results, but %
the $E_p$-dependence of $P$ in $H_{\perp c,E_p}=5$ T shown in Fig. \ref{E_vs_P} suggests that the third scenario %
contributes to the reduction of $P$. %

In Fig. \ref{FC_Pyro}(a-1), we show the temperature variations of $P$ under $H_{\parallel c}$. %
In contrast to the in-plane magnetic fields, the application of $H_{\parallel c}$ enhances $P$. %
Moreover, Fig. \ref{E_vs_P} shows that the sensitivity of $P$ to $E_p$ was significantly enhanced by applying $H_{\parallel c}$. %
As mentioned in introduction, the previous study on CFAO and CFGO %
has revealed that the sensitivity of $P$ to $E_p$ is determined by the `inhomogeneity' of the FE-ICM order, %
that must be relevant to the mobility of the magnetic domain walls in the FE-ICM phase.\cite{CompHelicity} %
Taking account of the present neutron diffraction measurements revealing that CFAO($x=0.015$) exhibits the long-range-ordered %
FE-ICM phase above $H_{\parallel c}=4$ T, %
we attributed to the enhanced sensitivity of $P$ to $E_p$ to the %
$H_{\parallel c}$-dependence of  the magnetic correlation in the FE-ICM phase. %

Note that the curvatures of the $E_p$ dependence of $P$ implies that the saturation value of $P$ was also enhanced by $H_{\parallel c}$. %
This might be ascribed to the $H_{\parallel c}$-dependence of the (local) magnetic structure including the wave number of the magnetic %
order in the FE-ICM phase. %

\begin{figure}[t]
\begin{center}
	\includegraphics[clip,keepaspectratio,width=7.5cm]{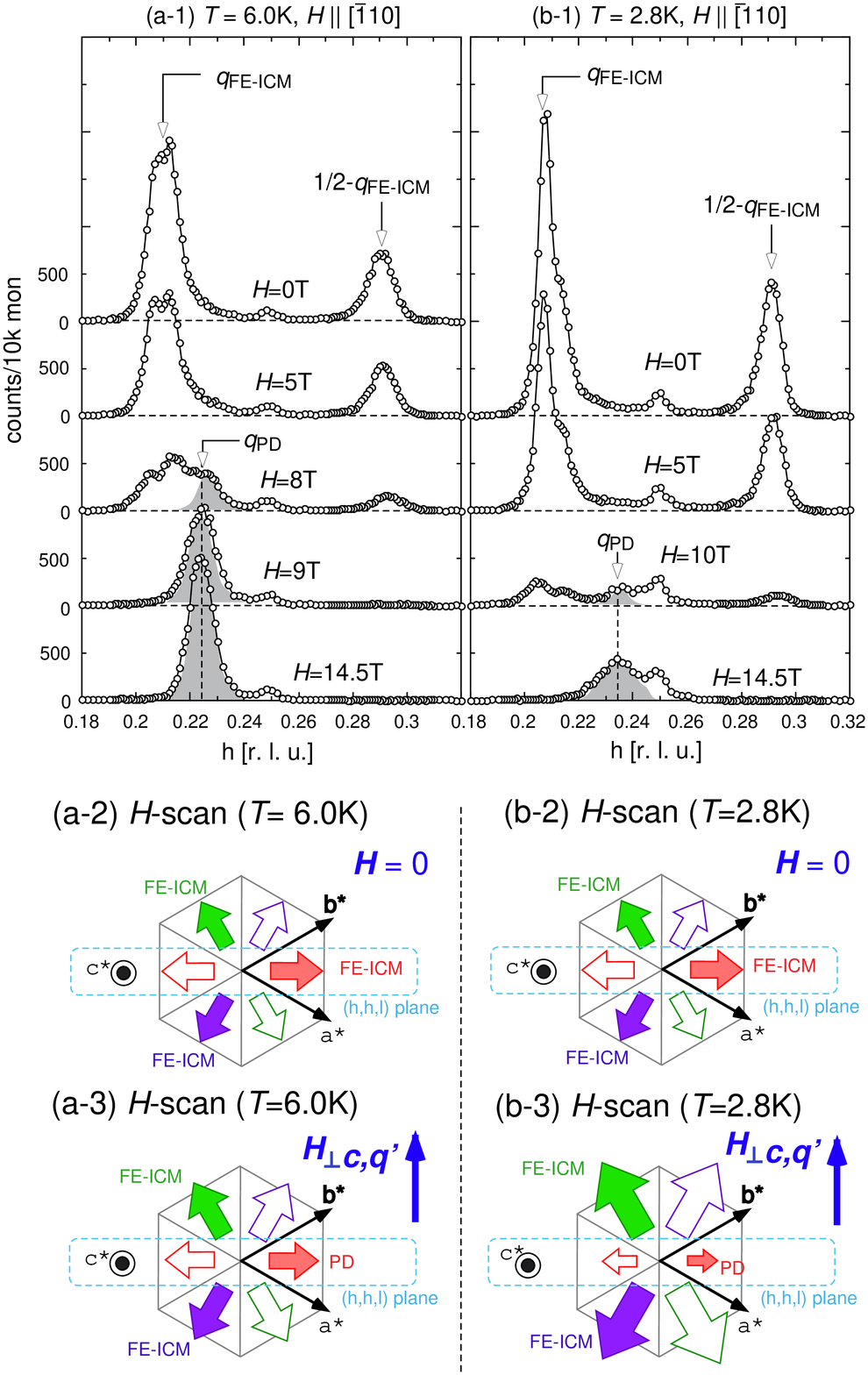}
	\caption{[(a),(b)] The $H_{\perp c,q'}$-variations of the neutron diffraction profiles of the $(h,h,\frac{3}{2})$ reciprocal lattice scans %
	at (a) 6.0 K and (b) 2.8 K. %
	[(a-2),(a-3),(b-2),(b-3)] Schematic drawings of the $H_{\perp c,q'}$-induced magnetic phase transitions in each $q$-domain. %
	The sizes of the arrows qualitatively show the volume fractions of the $q$-domains. %
	}
	\label{Neutron_H1-10}
\end{center}
\end{figure}

\subsubsection{Neutron diffraction measurements under applied magnetic field}

On the basis of the results of the field-cooling scans, %
we now discuss magnetic field variations of the FE-ICM order including those of the domain structure. %

According to the $H_{\perp c,q'}$-$T$ phase diagram shown in Fig. \ref{FC_PhaseDiagram}(c), %
the magnetic phase transition from the noncollinear FE-ICM phase to the collinear PD phase is expected %
in a $H_{\perp c,q'}$-increasing process. %
Figures \ref{Neutron_H1-10}(a-1) and \ref{Neutron_H1-10}(b-1) show the $H_{\perp c,q'}$-variations of the neutron diffraction %
profiles measured at $T=6.0$ K and 2.8 K, respectively. %
In the magnetic field scan at $T=6.0$ K, the intensities of the magnetic reflections corresponding to the %
FE-ICM order monotonically decrease with increasing $H_{\perp c,q'}$, and disappear around 9.0 T, %
as expected from the phase diagram. %
On the other hand, magnetic reflections described by the wave vector of the %
PD phase, $(q_{\rm PD},q_{\rm PD},\frac{3}{2})$ with $q_{\rm PD}\sim 0.22$, emerge above 8.0 T. %
This result apparently manifests the magnetic field induced phase transition from the FE-ICM phase to the PD phase. %

\begin{figure}[t]
\begin{center}
	\includegraphics[clip,keepaspectratio,width=7.5cm]{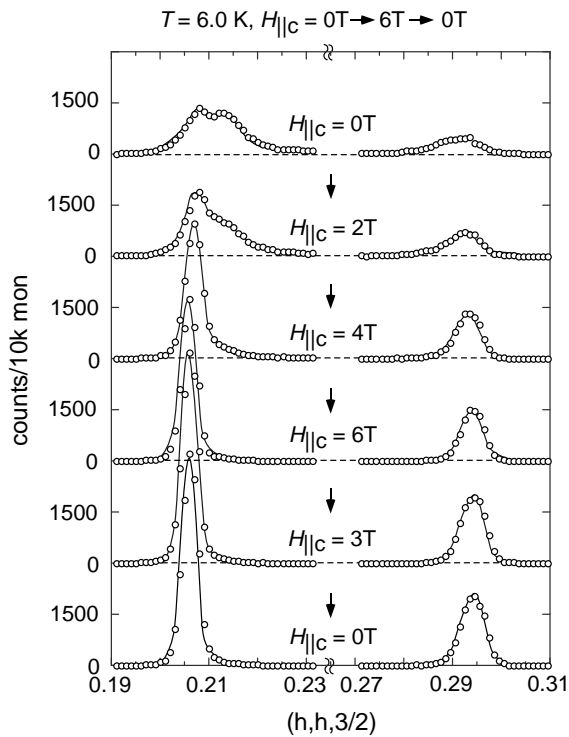}
	\caption{The $H_{\parallel c}$-variation of the neutron diffraction profiles of the $(h,h,\frac{3}{2})$ reciprocal lattice scans at $T=6.0$ K. %
	}
	\label{Neutron_H001}
\end{center}
\end{figure}

\subsection{Magnetic field variations of the FE-ICM order}
\label{Sec_MEeffect}

\begin{figure*}[t]
\begin{center}
	\includegraphics[clip,keepaspectratio,width=0.95\textwidth]{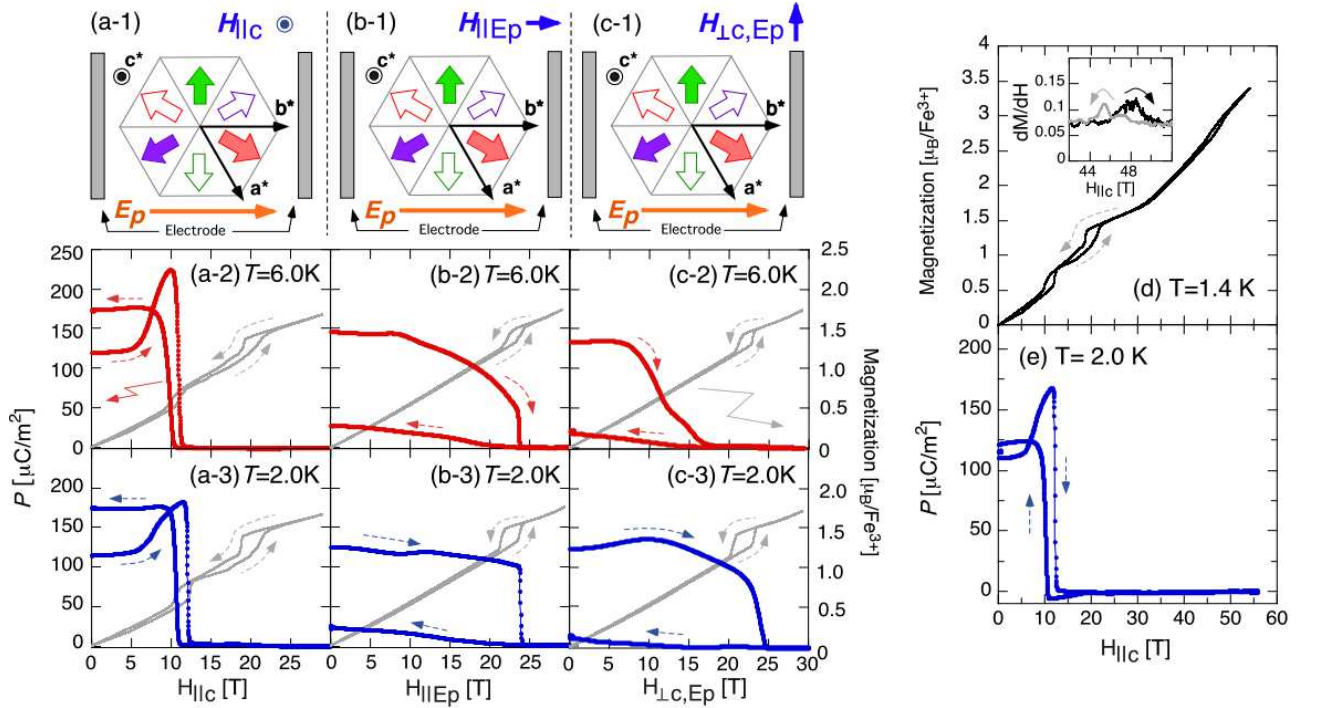}
	\caption{[(a-1),(b-1),(c-1)] The relationships between the directions of $E_p$, the $q'$-vectors and the applied magnetic fields of %
	(a-1) $H_{\parallel c}$, (b-1) $H_{\parallel E_p}$ and (c-1) $H_{\perp c,E_p}$. %
	The results of the polarization and magnetization measurements under the pulsed magnetic fields of [(a-2),(a-3)] $H_{\parallel c}$, %
	[(b-2),(b-3)] $H_{\parallel E_p}$ and [(c-2),(c-3)] $H_{\perp c,E_p}$. %
	[(d)-(e)] The  $H_{\parallel c}$-dependence of the magnetization and $P$ in CFAO($x=0.015$) up to 55 T. The inset shows the %
	$H_{\parallel c}$-dependence of $dM/dH_{\parallel c}$ around the fifth-field-induced phase transition. }
	\label{Pulse_P_M}
\end{center}
\end{figure*}

In the magnetic field scan at $T=2.8$ K, the field induced magnetic transition was also observed, as shown %
in Fig. \ref{Neutron_H1-10}(b-1). %
However, we found that the diffraction profile of the field-induced PD phase is rather diffusive. %
This indicates that the sinusoidally amplitude modulated magnetic structure of the PD phase is no longer stable at low temperatures. %
In addition, we also found that the intensity corresponding to the field induced PD magnetic order is considerably small. %
This implies that the volume fraction of the $(q,q,\frac{3}{2})$-domains is reduced by the applied magnetic field, % 
namely that the volume fractions of the three $q$-domains are changed by $H_{\perp c,q'}$, %
as illustrated in Figs. \ref{Neutron_H1-10}(b-2) and \ref{Neutron_H1-10}(b-3). %
This indicates that at low temperature, the in-plane magnetic field favors the proper helical magnetic domains whose helical axis follows it, %
as demonstrated in the previous works.\cite{CFAO_2q,Seki_CFGO} %

To summarize the $H_{\perp c,q'}$-dependence of the magnetic ordering in this system, %
the magnetic phase transition from the FE-ICM phase to the PD phase is observed at the relatively high temperature ($T=6.0$ K), %
and the re-population of the $q$-domains occurs at low temperatures ($T=2.8$ K). %
As mentioned in introduction, Seki \textit{et al.} have demonstrated that a magnetic field rotating in the triangular lattice plane %
can induce 120$^{\circ}$-flop of the electric polarization because of the re-population of the $q$-domains.\cite{Seki_CFGO} %
The maximum magnitude of the rotating magnetic field shown in Ref. \onlinecite{Seki_CFGO} was 6.5 T, and the measurements %
were carried out at 2.0 K. %
However, the present results suggest that more diversity in the ME-responses should be found by applying magnetic fields beyond $\sim 9$ T and %
by changing the temperature. %

Figure \ref{Neutron_H001} shows the $H_{\parallel c}$-variation of the neutron diffraction profiles at $T=6.0$ K. %
As seen in the field-cooling scans, the magnetic field applied along the $c$ axis sharpens the magnetic diffraction profile. %
In addition, the wave number of the FE-ICM order, which was distributed around $q\sim0.21$ in zero field, was concentrated at $q=0.207$ %
above $H_{\parallel c}=4$ T. %
We also found that the sharp magnetic diffraction profiles retained after removing the magnetic field. %
This history dependent behavior indicates the long-range-magnetic ordering realized by the application of $H_{\parallel c}$ remained even %
after returning to  the zero field. %

\subsubsection{magnetization and dielectric polarization measurements under pulsed magnetic fields}
\label{Pulse}
We also performed the dielectric polarization and magnetization measurements %
under pulsed magnetic field up to 30 T. %
The electrode configuration and the magnetic field directions were selected to be the same as those in the pyroelectric measurements discussed in Sec. \ref{Sec_Pyro}. %

We firstly discuss the metamagnetic transition in this system using the results of the magnetization measurements. %
In $H_{\parallel c}$ up to 30 T, three magnetic phases appear, as shown in Figs. \ref{Pulse_P_M}(a-2)-\ref{Pulse_P_M}(a-3). %
In the previous work on CFAO($x=0.02$),\cite{Terada_CFAO_H-T} %
Terada \textit{et al.} have reported that the first field induced phase from the FE-ICM phase %
is the slightly incommensurate magnetic phase referred to as the `FI phase'. %
We thus considered that the first field induced phase from the FE-ICM phase in CFAO ($x=0.015$) is the FI phase. %
Around $H_{\parallel c}=20$ T, where the induced magnetization approaches $\sim 5/3 \mu_{\rm B}$, the system undergoes %
further magnetic phase transition. %
In the in-plane magnetic fields, magnetization plateaus with the magnetization of $\sim 5/3 \mu_{\rm B}$ %
were also observed around 23 T. %
Comparing the results of the previous magnetization measurements on pure CFO under pulsed magnetic field,\cite{Terada_CFO_Pulse} %
we realize that this magnetic phase is the 3-sublattice (3SL) phase, %
which is also observed in pure CFO.\cite{Terada_CFO_Pulse} %

Let us move on to the results of the dielectric polarization measurements. %
In $H_{\parallel c}$, the finite electric polarization was observed only in the FE-ICM phase, as shown in Figs. \ref{Pulse_P_M}(a-2)-%
\ref{Pulse_P_M}(a-3). %
In the $H_{\parallel c}$-increasing process, $P$ rapidly increases in the magnetic field region of $5<H_{\parallel c}<12$ T. %
This enhancement corresponds to the magnetic field dependences of the magnetic correlation and the magnetic modulation wave number, %
which were observed in the present neutron diffraction measurements. %
In the $H_{\parallel c}$-decreasing process, %
$P$ is also observed to emerge in the FE-ICM phase. In contrast to the $H_{\parallel c}$-increasing process, %
the value of $P$ is almost independent of the magnetic field, and is larger than the value in the zero field state before the measurement. %
Judging from the symmetry of the magnetic and crystal structure in this system, it is reasonable to assume that the volume fractions of the three $q$-domains are not changed by $H_{\parallel c}$. %
Hence, we ascribed this $H_{\parallel c}$-variations of $P$ to the history dependent behavior of the FE-ICM order in a $q$-domain %
observed in the present neutron diffraction measurements. %

Figures \ref{Pulse_P_M}(b-3) and \ref{Pulse_P_M}(c-3) show the `in-plane' magnetic field dependences of $P$ measured at %
relatively low temperature, $T=2.0$ K. %
From the results of the present neutron diffraction measurements, %
it is expected that the magnetic-field-induced re-populations of the $q$-domains result in anisotropic magnetic field variations of $P$. %
However, the expected anisotropic behaviors are not observed, %
although the details of the $P$-$H_{\perp c,E_p}$ and $P$-$H_{\parallel E_p}$ curves are slightly different from %
each other. %
This result suggests that the sweeping rate of the pulsed magnetic fields were too fast to induce the re-population of the $q$-domains. %

On the other hand, at $T=6.0$ K, we found that the in-plane field variations of $P$ were quite anisotropic, as shown in Figs. \ref{Pulse_P_M}(b-2) and %
\ref{Pulse_P_M}(c-2). %
In the $H_{\perp c,E_p}$-increasing process, the value of $P$ started to decrease around 8 T, and disappeared around 15 T, which is far below %
the transition field to the 3SL phase. %
In contrast,  in the $H_{\parallel E_p}$-increasing process, the finite value of $P$ is observed up to the transition field to the 3SL phase. %
These results revealed that %
the `in-plane' field induced FE-ICM to PD transition %
in a $q$-domain occurs even in the pulsed magnetic fields. %
It should be noted that this field induced FE-ICM to PD phase transition is not clearly observed in the magnetization %
measurement. %
This is because the FE-ICM to PD phase transition occurs only in the $q$-domains with the wave vectors %
of $(q,q,\frac{3}{2})$ and $(-q,2q,\frac{3}{2})$ while all the three $q$-domains contributes to the measured magnetization. %

We also found that in the $H_{\perp c,E_p}$- and $H_{\parallel E_p}$-decreasing process, the values of $P$ are %
significantly smaller compared to the values in zero field before the measurements, in contrast to the $H_{\parallel c}$-decreasing %
process, in which $P$ was comparable to the value at the initial state. %
At this stage, we have no clear explanation for this result. %
In order to investigate this point, further neutron diffraction measurements under applied magnetic fields are required. %

In Figs. \ref{Pulse_P_M}(d) and \ref{Pulse_P_M}(e), we show the $H_{\parallel c}$-dependence of the magnetization and $P$ up to 55 T, %
respectively. %
Although, in this paper, we do not focus on the high-magnetic-field behavior in this system, %
it is worth mentioning here that the fifth-magnetic-field-induced phase transition, which has been recently found in %
pure CFO,\cite{CFO_HighField,CFO_HighField2} was also observed in CFAO($x=0.015$) around $H=48$ T. %
No finite electric polarization was detected in the magnetic-field-induced phases including the `fifth' phase. %

\section{CONCLUSION}
\label{Conc}

We have investigated the magnetic field dependence of the FE-ICM order in the trigonal ME-multiferroic %
CuFe$_{1-x}$Al$_x$O$_2$ with $x=0.015$ by means of the neutron diffraction, dielectric polarization and magnetization measurements %
under the magnetic fields applied along various directions. %

We have established the $H$-$T$ magnetic %
phase diagrams for the three principal directions of the magnetic fields, $H_{\parallel c}$, $H_{\parallel q'}$ and $H_{\perp c,q'}$. %
It should be emphasized that these $H$-$T$ phase diagrams represent the magnetic ordering `in a $q$-domain', %
and reveal the anisotropic in-plane-magnetic-field responses of the FE-ICM order in a $q$-domain, %
which was not directly observed in the previous macroscopic (bulk) polarization and magnetization measurements.\cite{Seki_CFGO} %

We have also found that the sensitivity of $P$ to $E_p$ in this system was controlled by applying a magnetic field. %
This indicates that the `inhomogeneity' of the FE-ICM order, which is also controlled by an applied magnetic field, is relevant to the %
sensitivity of $P$ to $E_p$. %
This result is consistent with our previous polarized neutron diffraction study on CFAO($x=0.015$) and CFGO($x=0.035$).\cite{CompHelicity} %

While the recent dielectric polarization measurements on CuFe$_{1-x}$Ga$_x$O$_2$ with $x=0.035$ by Seki \textit{et al.} %
have demonstrated that %
the magnetic field dependence of the `magnetic domain structure', specifically the volume fractions of the three $q$-domains, %
results in the distinct magnetic field responses of $P$,\cite{Seki_CFGO} %
the present results have revealed the anisotropic magnetic field dependence of the FE-ICM order `in each $q$-domain' %
can be also a source of a variety of the magnetic field dependence of $P$ in this system. %

\section*{Acknowledgments}
The neutron diffraction measurement at BENSC was carried out along the proposals PHY-01-2285-DT and PHY-01-1878. %
This work was supported by a Grant-in-Aid for Scientific Research (C), No. 19540377, and a Grant-in-Aid for JSPS research fellow, No. 21.2745 %
from JSPS, Japan. %
This work was also supported in part by a Grant-in-Aid for Scientific Research on priority Areas `High Field Spin Science in 100T' (No.451) from the Ministry of Education, Culture, Sports, Science and Technology (MEXT). %
The images of the crystal and magnetic structures in this paper were depicted using the software VESTA\cite{VESTA}  %
developed by K. Monma.%


\begin{thebibliography}{99}
\bibitem{comment2} Although the magnetic field was not exactly parallel to the $c$ axis as mentioned in Sec. \ref{Experimental_details}, %
we consider that the in-plane component of the magnetic field hardly affected the magnetic ordering in this system, %
because the direction of the in-plane component of the magnetic field is parallel to the $q$'-vector of the $(q,q,\frac{3}{2})$-domain. %
\bibitem{comment1} Note that the two reflections in the FE-ICM phase should be properly assigned as $(0,q,\frac{1}{2})$ and $(0,1-q,\frac{1}{2})$ using the monoclinic basis employed in the previous works. %
However, in this paper, we employed the conventional hexagonal basis in order to describe the directions of the magnetic modulation wave vectors of the three $q$-domains. %
\end{thebibliography}
\end{document}